\documentclass[fleqn,usenatbib]{mnras}

\usepackage[T1]{fontenc}
\usepackage{lscape}

\DeclareRobustCommand{\VAN}[3]{#2}
\let\VANthebibliography\thebibliography
\def\thebibliography{\DeclareRobustCommand{\VAN}[3]{##3}\VANthebibliography}

\usepackage{graphicx} 
\usepackage{amsmath} 
\usepackage{amssymb} 
\usepackage[normalem]{ulem}
\usepackage{tablefootnote}

\title[Spectroscopic diagnostics of coronal fan loops]{ Spectroscopic diagnostics of coronal fan loops using EIS/Hinode and AIA/SDO}

\author[Rawat \& Gupta]{Ananya Rawat$^{1}$\thanks{E-mail: ananyarawat@prl.res.in (AR)}, Girjesh Gupta$^{1}$
\\
$^{1}$Udaipur Solar Observatory, Physical Research Laboratory, Dewali, Badi Road, Udaipur 313001, India \\
}

\date{Accepted . Received ; in original form}

\pubyear{2026}
\begin{document}
\label{firstpage}
\pagerange{\pageref{firstpage}--\pageref{lastpage}}
\maketitle

\begin{abstract}

We utilize spectroscopic and imaging data from the EIS/Hinode and the AIA/SDO to determine plasma and wave parameters along coronal fan loops. We determined the temperature along the loop lengths using the EM-loci method with EIS spectral lines and identified multi-thermal plasma components at temperatures ranging from 0.8-1.05, 1.2-1.4, and 1.6-1.8 MK. We further compared these temperature distributions along the loops with those derived from DEM analysis carried out using AIA coronal passbands. The multi-thermal nature of coronal fan loops can be well resolved using EIS spectroscopic data. 
We determined the density along the loop using the line-ratio method from the EIS Fe XII line pair. We found that density decreases slowly from the expected hydrostatic equilibrium. Our results demonstrate the multi-thermal nature of overdense loops, which can be well resolved with EIS spectroscopic data. We also determined the velocity amplitude of Alfv\'en waves from the non-thermal broadening of EIS Fe XII spectral line profiles. We further estimated the Alfv\'en wave energy flux in the corona by combining these parameters with the recently reported magnetic field strength of $\approx$200 G at the coronal footpoint of fan loops. The estimated Alfvén wave energy flux is $\approx$ 2$\times 10^7$ erg cm$^{-2}$ s$^{-1}$, significantly higher than those reported in earlier studies, and can contribute to coronal losses due to radiation along the fan loops in the active region.

\end{abstract}


\section{Introduction}
\label{sec:intro} 

The high temperature of the solar corona is 80-year-long-standing problem in solar physics. Magnetic loops are considered the building blocks of the solar atmosphere, act as a waveguide and carry waves that may heat the solar atmosphere \citep{2014LRSP...11....4R,2023MNRAS.525.4815R,2024MNRAS.533.1166R}. These loops are classified into three categories depending on their characteristic temperatures: cool loops ($\leq$ 1 MK), warm loops ($\approx$ around 1 MK), and hot loops (3-5 MK) \citep{2014LRSP...11....4R,2015ApJ...800..140G}. Loops are distinguishable in corona as their intensity is about 20\%–30\% stronger than foreground/background diffuse emission \citep{2009ApJ...694.1256T,2012ApJ...753...35V}. To calculate various plasma parameters along the loops \citep[e.g.,][]{2009ApJ...694.1256T,2015ApJ...800..140G,2019A&A...627A..62G}, various spectroscopic and imaging data techniques are developed, see details in the \citet{2008uxss.book.....P}. 

The ubiquitous presence of outward propagation and damping of MHD waves along coronal loops are reported in the solar corona \citep[e.g.,][]{2007Sci...317.1192T,2017ApJ...836....4G,2026ApJR}. These waves are reported to originate at the solar photosphere \citep{2023MNRAS.525.4815R,2025ApJ...988L..26R} and show evidence of damping while propagating \citep{2024MNRAS.533.1166R}. Alfv\'en waves are considered a potential candidate for heating the solar corona, as they can carry a large energy flux generated at the photosphere along magnetic field lines. However, it is difficult to explain how they can be a major supplier of heat to coronal plasma as the signature of dissipation of these wave energies into heat energy is still not observed \citep[e.g.,][]{1997ApJ...488..895H, 2011ApJ...736....3V}. Therefore, Alfv\'en waves require an efficient dissipation mechanism to explain their role in plasma heating. The various proposed Alfv\'en wave damping mechanisms are phase mixing, mode conversion, resonant absorption, etc. \citep[see the recent review of][and references therein]{2020SSRv..216..140V}. 

\begin{figure*}
    \centering
    \includegraphics[width=\linewidth]{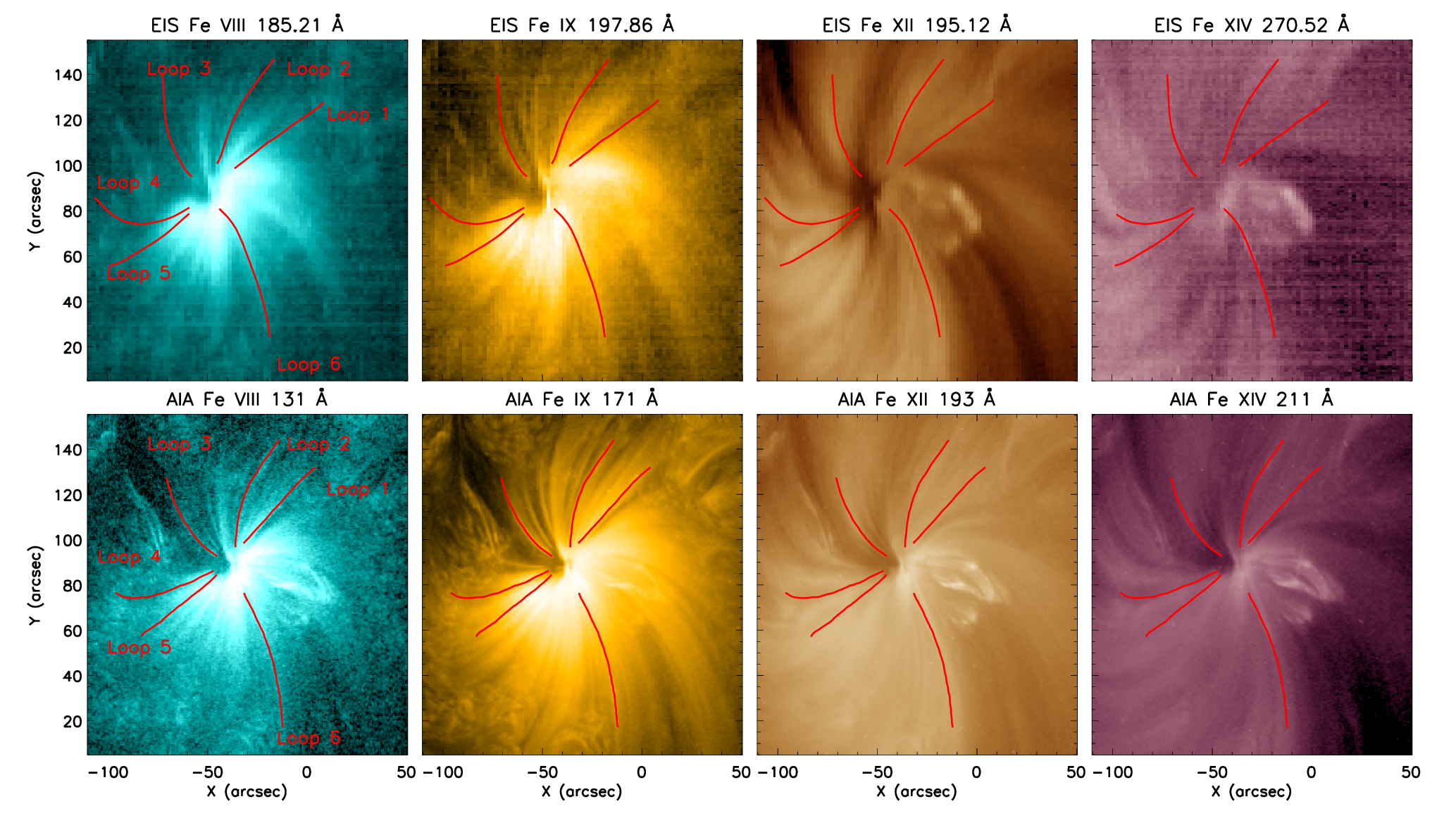}
    \caption{Top panels: Monochromatic intensity map of fan loop system obtained from various EIS spectral lines as labeled. Bottom panels: Images of the fan loop system from the different AIA passbands sensitive to the same ions as in the top panels.}
    \label{fig:fan loop}
\end{figure*}    

To accurately estimate wave energy flux in the solar atmosphere, remote sensing of the local plasma parameters is of major importance. Accurate diagnostics of plasma parameters in terms of density, velocity, temperature, and composition are crucial for understanding local physical processes. 
Various imaging and spectroscopic diagnostic techniques have been developed to determine the plasma parameters. For imaging data, the differential emission measure \citep[DEM;][]{1972SoPh...22..307J} technique is utilized to quantify the amount of plasma present at a given temperature. Plasma density and temperature measurements rely on accurate knowledge of various atomic parameters, provided by the CHIANTI database \citep{1997A&AS..125..149D} and the response functions of each instrument. Recently, a new technique, RFit (Response Fitting), has been introduced, which provides the relative percentage contributions of the hot-cool emission from different AIA passbands \citep{2024SoPh..299...94A,2026SoPh..301...70A}.

For spectroscopic data, spectral line analysis can provide constraints on density and temperature because spectral profiles depend on these parameters, particularly through collisional excitation/de-excitation and level populations. There are conflicting opinions on the multi-thermal temperature distribution due to inconsistent measurements from different instruments and techniques, as reported in a few studies \citep[e.g.,][]{2011ApJ...731...49S,2011ApJ...730...85B}.

Upon using average coronal magnetic field strength of $\approx$ 10-50 G in coronal active regions \citep[e.g.,][]{2000ApJ...541L..83L,2008A&A...487L..17V}, the reported Alfv\'en wave energy fluxes were of the order $\approx 10^6$ erg cm$^{-2}$ s$^{-1}$ \citep[e.g.,][]{2007Sci...317.1192T,2019A&A...627A..62G}. These wave energy fluxes are less than the energy flux required to maintain the hot active region corona, i.e., $\approx 10^7$  erg cm$^{-2}$ s$^{-1}$\citep{1977ARA&A..15..363W}. Therefore, it was considered that although Alfv\'en waves can heat the quiet Sun corona, they can not heat the hot and dense active region corona \citep[e.g.,][]{2007Sci...317.1192T,2011Natur.475..477M}. However, in recent works, \citet{2021ApJ...915L..24B,2025ApJ...988L..26R} have reported significantly higher magnetic field strengths and Alfv\'en wave speeds along coronal loops. These recent measurements can significantly enhance the previously estimated Alfv\'en wave energy fluxes. 

For accurate estimation of plasma and wave parameters, it is mandatory to have excellent signals along stable loops, which is not always the case. In this paper, we present a comparison of plasma parameters obtained from imaging and spectroscopic instruments to explore plasma and wave properties along coronal fan loop structures. For this purpose, we identified a unique set of spectroscopic data scanning a fan loop system with a 1" slit, providing better spatial resolution. These structures provide an excellent site to study plasma parameters and wave properties in the corona. Details of the observations are described in Section 2. We employ several imaging and spectroscopic diagnostics to estimate the temperature, density, and non-thermal velocity amplitude along these loops, which are described in Section 3. We further estimate the Alfv\'en wave energy flux along coronal fan loops. The obtained results are discussed and summarized in Section 4. \\


\section{Observations}
\label{sec:obs}

To investigate plasma and waves properties along coronal loops, we selected a fan loop system belonging to active region NOAA AR 11101, as shown in \autoref{fig:fan loop}. The fan loop system was observed simultaneously by Extreme Ultraviolet Imaging spectrometer \citep[EIS;][]{2007SoPh..243...19C} onboard \citep[Hinode;][]{2007SoPh..243....3K} and  Atmospheric Imaging Assembly \citep[AIA;][]{2012SoPh..275...17L} onboard Solar Dynamics Observatory \citep[SDO;][]{2012SoPh..275....3P} on 2010 August 30. We obtained 2 h of data starting from 11:13:30 UT. AIA extreme ultraviolet (EUV) images provide good coverage over the transition region and the corona at a resolution of 0.6"/pixels and a cadence of 12 s. 

The EIS raster scan used a 1" slit, 60 s exposure time, and 2" step size, covering a 240" $\times$ 512" FOV in 120 min, starting from 11:13:34 UT. Observations were performed over the wavelength range of 180-204 \AA\ and 248-284 \AA. We followed standard procedures to prepare the EIS data using the IDL routine EIS$\_$PREP, available in the Solar Software \citep[SSW;][]{1998SoPh..182..497F}, which calibrates and corrects for slit tilt and orbital drift. The spatial offset in the solar Y-direction at different wavelengths is corrected by co-aligning the images with EIS Fe XII 195.12 \AA\, the strongest line in our analysis. 
  
Imaging data allows us to co-align data from different instruments, enabling the simultaneous use of data from different instruments operating at different wavelengths. We co-aligned EIS and AIA observations using EIS Fe XII 195 \AA\ and AIA 193 \AA\ images, respectively, using the cross-correlation method. All AIA images are derotated with respect to the EIS starting observation time, i.e., 11:13 UT. We have fitted all the EIS spectral line profiles with a Gaussian function using EIS$\_$AUTO$\_$FIT. This routine provides 1$\sigma$ error on the fitted parameters. Since EIS sensitivity is deteriorating, we have further recalibrated the obtained intensity and error bars using \citet{2013A&A...555A..47D}.  In the top panel of \autoref{fig:fan loop}, we plot the monochromatic intensity map of the fan loop system obtained from various EIS spectral lines. In the bottom panel of \autoref{fig:fan loop}, we plot AIA images sensitive to the same spectral line of EIS. 

 
The identified data set provides a unique opportunity to study the spatial, temporal, and spectral evolution of waves in intensity and Doppler broadening along the plane-of-sky (POS) and line-of-sight (LOS), respectively, in the corona. Additionally, this is among the few observations when the fan loops emanating from a unipolar sunspot near the disk center are observed with a 1" slit. 

\section{Analysis and Results}
\label{sec:analysis}

\autoref{fig:fan loop} shows the analyzed fan loop structure in the solar corona observed by AIA and EIS as labeled. We have identified six fan loops that are clearly visible in the hot and cool passbands and in spectral lines, and can be traced in both AIA and EIS images. The overplotted lines represent the manual tracing of the loops. The loops are traced separately in AIA and EIS images; therefore, there can be a slight offset along the loop locations obtained from AIA and EIS. In this work, we analyze all the identified loops. However, here we present results from Loop 1 as a representative example.  

\begin{figure*}
    \centering
    \includegraphics[width=0.99\linewidth]{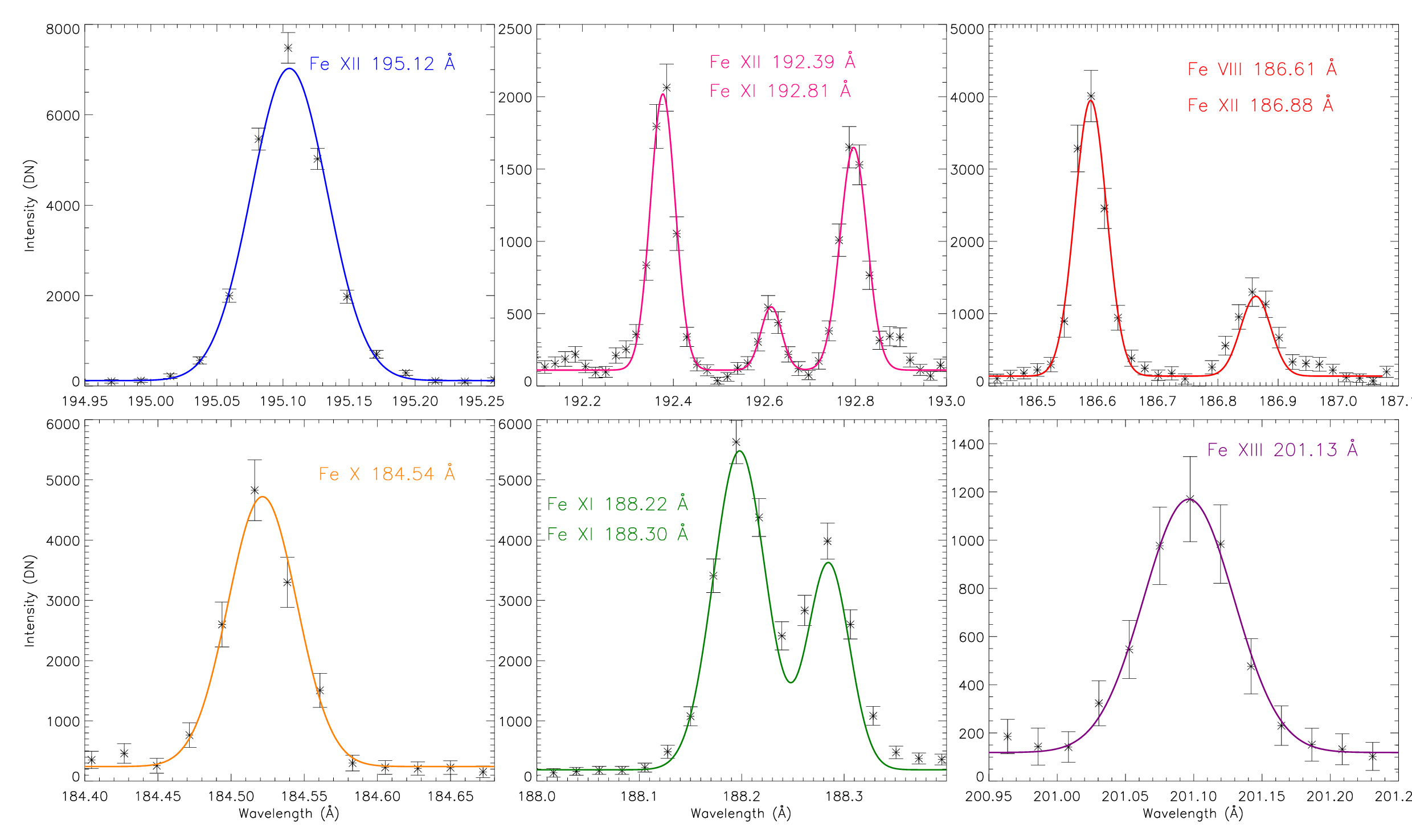}
    \caption{Spectral line profiles obtained at the coronal loop foot-point and fitted with the Gaussian functions.}
    \label{fig:prof}
\end{figure*}    

\begin{figure}
    \centering
    \includegraphics[width=0.99\linewidth]{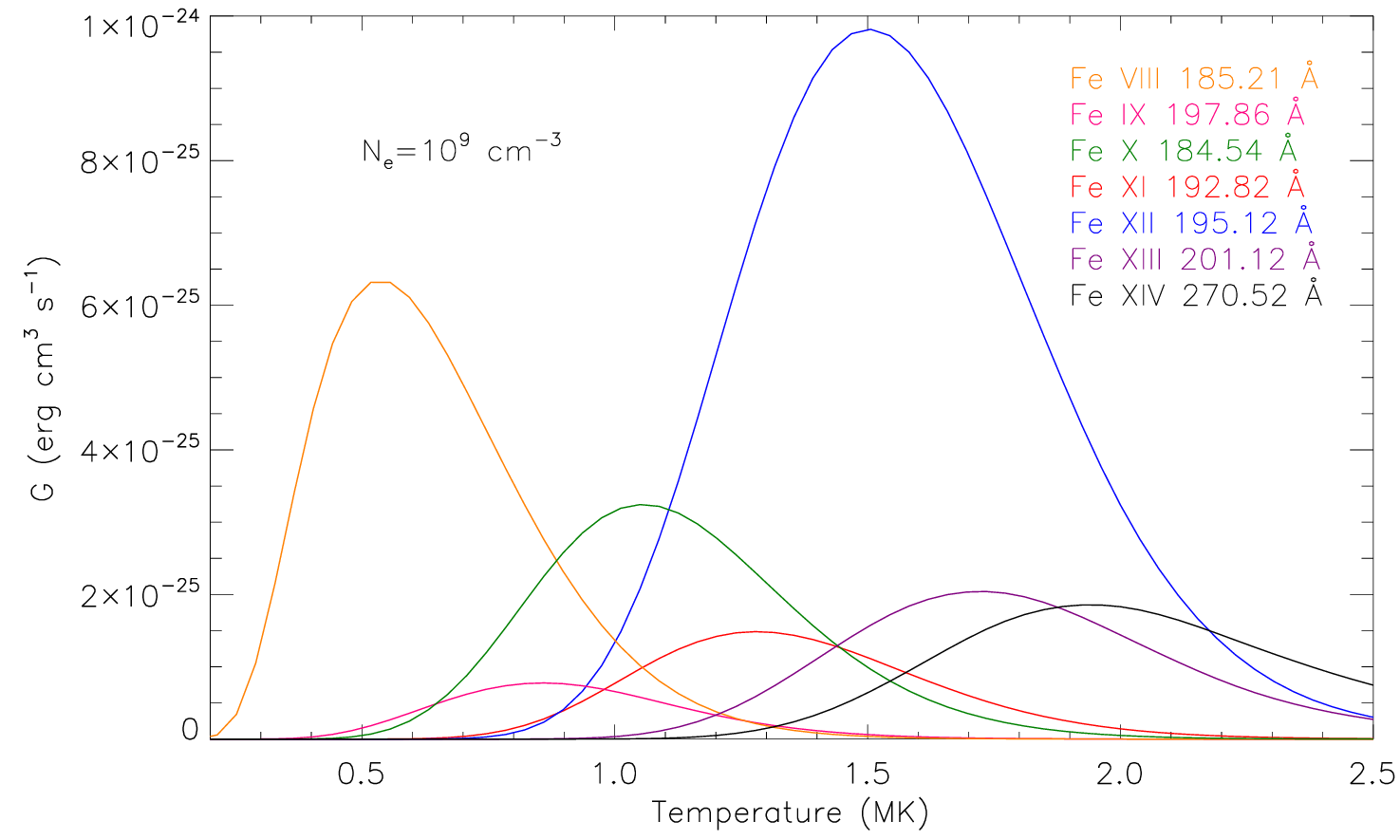}
    \caption{Contribution functions of various spectral lines used in this study.}
    \label{fig:cf}
\end{figure}

\subsection{Plasma diagnostics} 

For an optically thin corona, the emission line intensity (I) can be given as

\begin{equation}
    I (\lambda_{i,j})= \int C(T_e,N_e) N_e^2 dh,
\end{equation}
We also obtained the average emission measure (EM) variation along the loop from I using 

\begin{equation}
\label{eq:em}
    EM_{col} = \frac{I (\lambda_{i,j})}{C ( N_e , T )}= \int N_e^2 dh = \int_T DEM(T) dT.
\end{equation}
Here, the abundance is assumed to be constant over the source volume \citep{1992PhyS...46..202F}, $h$ is the column depth of emitting plasma along the LOS. $C(T_e, N_e )$ is the contribution function, which contains all the relevant atomic parameters for each spectral line. $T_e$ is electron temperature, $N_e$ is electron number density, i and j are the lower and upper levels, and $\lambda_{i,j}$ is the wavelength of the spectral line \citep{2008uxss.book.....P}. 
All the parameters within the contribution function are calculated using CHIANTI version 10.1 \citep{2021ApJ...909...38D} available in the SSW distribution at a constant electron number density $N_e= 10^9$ cm$^{-3}$. The contribution functions $C(T_e,N_e)$ of various spectral lines used for plasma diagnostics are plotted in \autoref{fig:cf}. DEM is the differential emission measure, which is a measure of the amount of plasma along the line-of-sight (LOS) that contributes to the emitted radiation at temperature $T_e$.

To perform spectral line analysis, we identified several isolated, unblended spectral lines with strong signals. The selected spectral lines provide good coverage over the fan loop temperature range 0.5-2 MK, as they are considered cool loops \citep[e.g.,][]{2017ApJ...835..244G}. The spectral lines used in this analysis are provided in \autoref{tab:line}. All the selected spectral line profiles are then fitted with a Gaussian function. \autoref{fig:prof} shows isolated spectral line profiles along with the fitted Gaussian profiles over them.

\begin{figure*}
    \centering
    \includegraphics[width=0.8\linewidth]{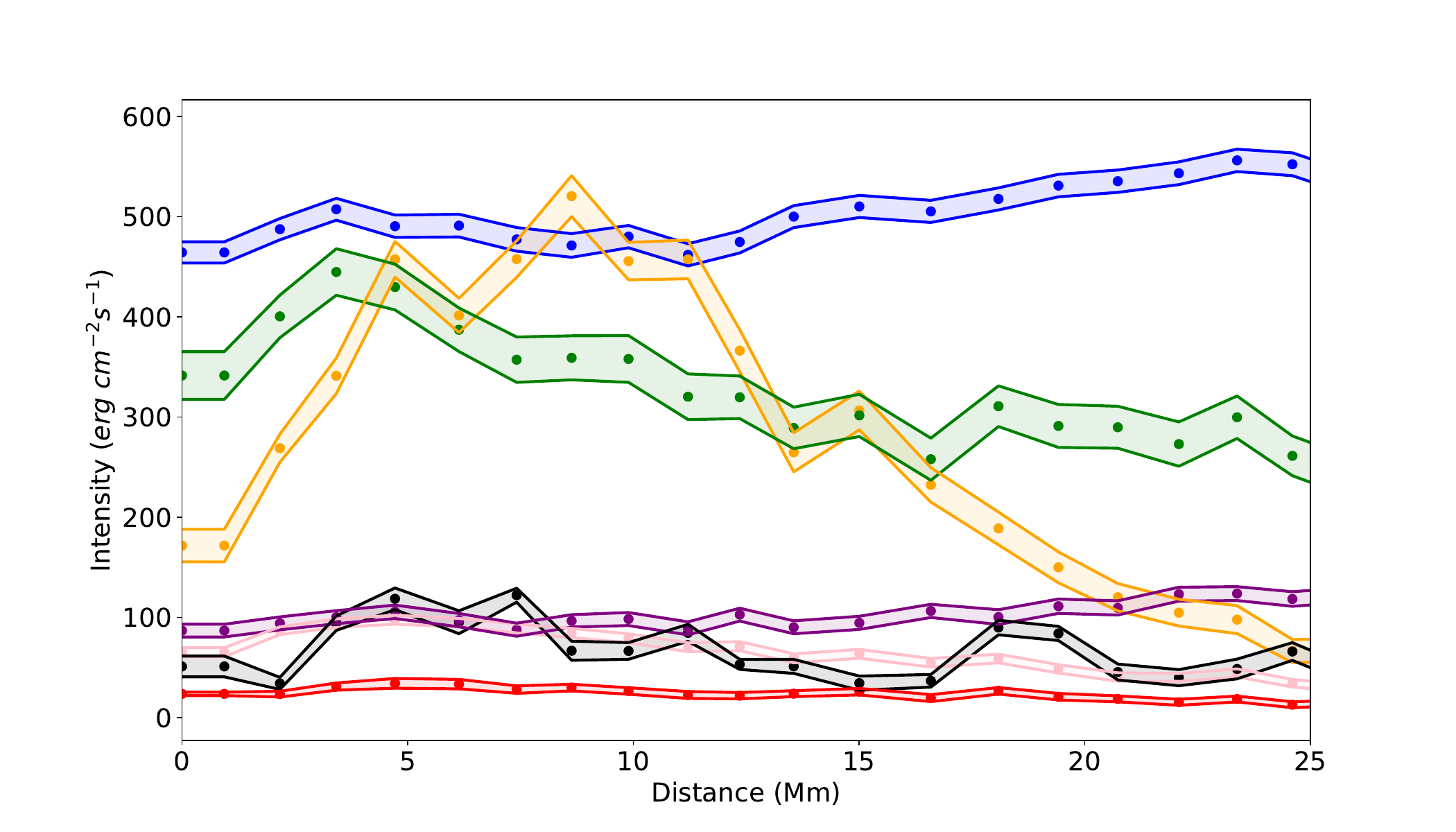}
    \caption{Intensity obtained from different spectral lines is plotted in different colors along the coronal loop. The color code for different ions is the same as \autoref{fig:cf}.}
    \label{fig:int}
\end{figure*}

In \autoref{fig:int}, we present the intensity variations along the loops for selected spectral lines, with the corresponding $1\sigma$ error bars. These intensities are further used to determine plasma parameters along the loop. Background subtraction plays an important role in determining plasma parameters along coronal loops \citep[e.g.,][]{2009ApJ...694.1256T}. We therefore attempted to estimate the background intensity in the vicinity of the loop. However, the active region exhibits structural variations between high- and low-temperature images, making it difficult to identify the loops and their corresponding background regions consistently. Therefore, due to the non-uniform background emission, we performed the plasma diagnostics using the original (unsubtracted) intensities. This approach may introduce LOS averaging effects into the derived plasma parameters.

\begin{table}
    \caption{Selected EIS spectral lines observed by different ions emitting different wavelengths along with their peak formation temperature. }
     \label{tab:line}
   \centering
    \begin{tabular}{|c|c|c|c|c|c|c|c|c|c|} 
    \hline  
\multicolumn{1}{|c|}{ Ion }  &
\multicolumn{1}{|c|}{ $\lambda_{obs}$ }  &
\multicolumn{1}{|c|}{Temp.}  \\ 

     & \AA\ & MK  \\ \hline
Mg V    &276.16  & 0.32 \\ \hline     
Fe VIII &185.21  & 0.54 \\ \hline
Fe IX   &197.86  & 0.86 \\ \hline
Fe X    &184.54  & 1.06 \\ \hline
Fe XI   &192.82  & 1.28 \\ \hline
Fe XII  &192.39  & 1.50 \\ \hline
Fe XII  &195.12  & 1.50 \\ \hline
Fe XII  &186.88  & 1.50 \\ \hline
Fe XIII &201.12  & 1.72 \\ \hline
Fe XIV  &270.53  & 1.94 \\ \hline
    \end{tabular}
\end{table}

\subsubsection{Temperature}

\begin{figure*}
    \centering
    \includegraphics[width=0.99\linewidth]{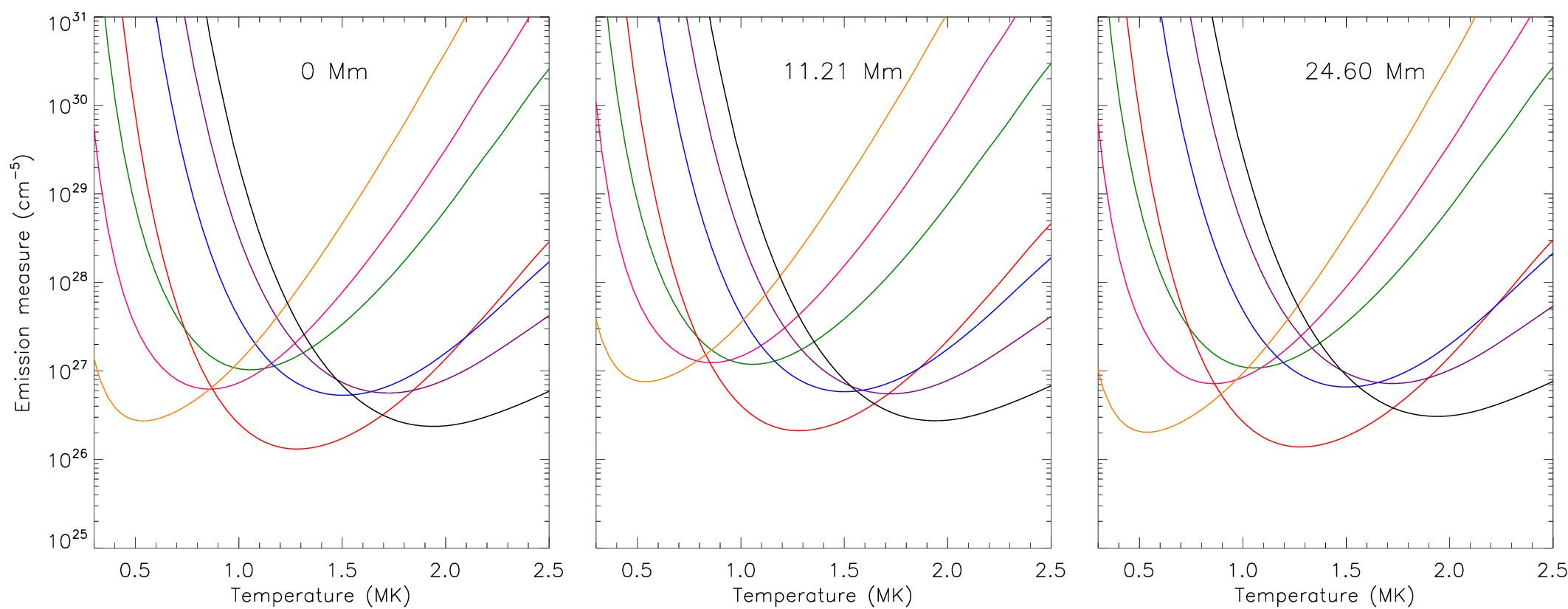}
    \caption{EM loci curves of spectral lines obtained from using \autoref{eq:em} at different loop lengths as labeled. The color code for different ions is the same as \autoref{fig:cf}.}
    \label{fig:em_loci}
\end{figure*}

The electron temperature ($T_e$) along the loops can be estimated using emission lines from ions of the same element at different ionization stages, sensitive to different temperature regimes. For this analysis, we used the density-insensitive emission lines Fe VIII 185.21 \AA, Fe IX 197.86 \AA, Fe X 184.53 \AA, Fe XI 192.82 \AA, Fe XII 192.39 \AA, Fe XIII 201.12 \AA\ and Fe XIV 270.53 \AA. To estimate the electron temperature, we utilize the emission measure (EM) loci method, following the approach described in \citet[e.g.,][]{2008uxss.book.....P}. In this method, we obtain the EM curves calculated using \autoref{eq:em}, which are ratios of observed emission line intensities (I) with the corresponding contribution functions (G) for different spectral lines. In \autoref{fig:em_loci}, we present the EM loci curves for three representative locations along the loops as labeled. The curves intersect at three temperature ranges rather than at a single temperature, suggesting that the loops exhibit a multi-thermal plasma structure.

\begin{figure*}
    \centering
    \includegraphics[width=0.8\linewidth]{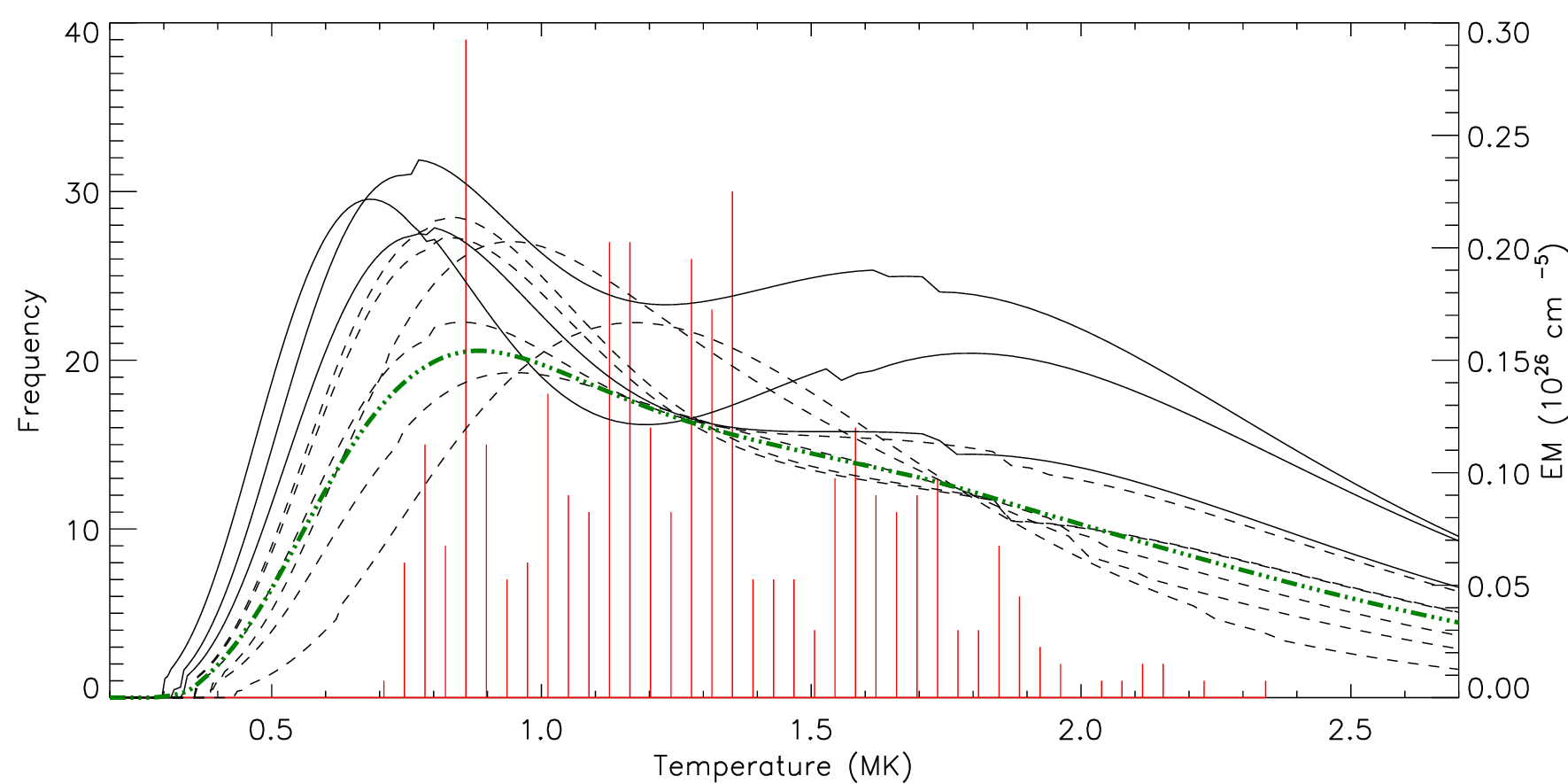}
    \caption{ The histogram of all the EM loci intersections along the loop. The x-axis represents temperature, and the left and right y-axis represents the frequency of intersections obtained from EM-loci method and emission measure obtained from DEM analysis respectively. The overplotted black lines show the EM curves obtained from DEM analysis along the loop, 5 pixels apart. The solid black lines show the DEM, which resolves the dual-temperature component, whereas the dashed black line shows single temperature component. The green dash-dot lines show the average EM profile along the loop.}
    \label{fig:histo}
\end{figure*}

\begin{figure*}
    \centering
    \includegraphics[width=0.95\linewidth]{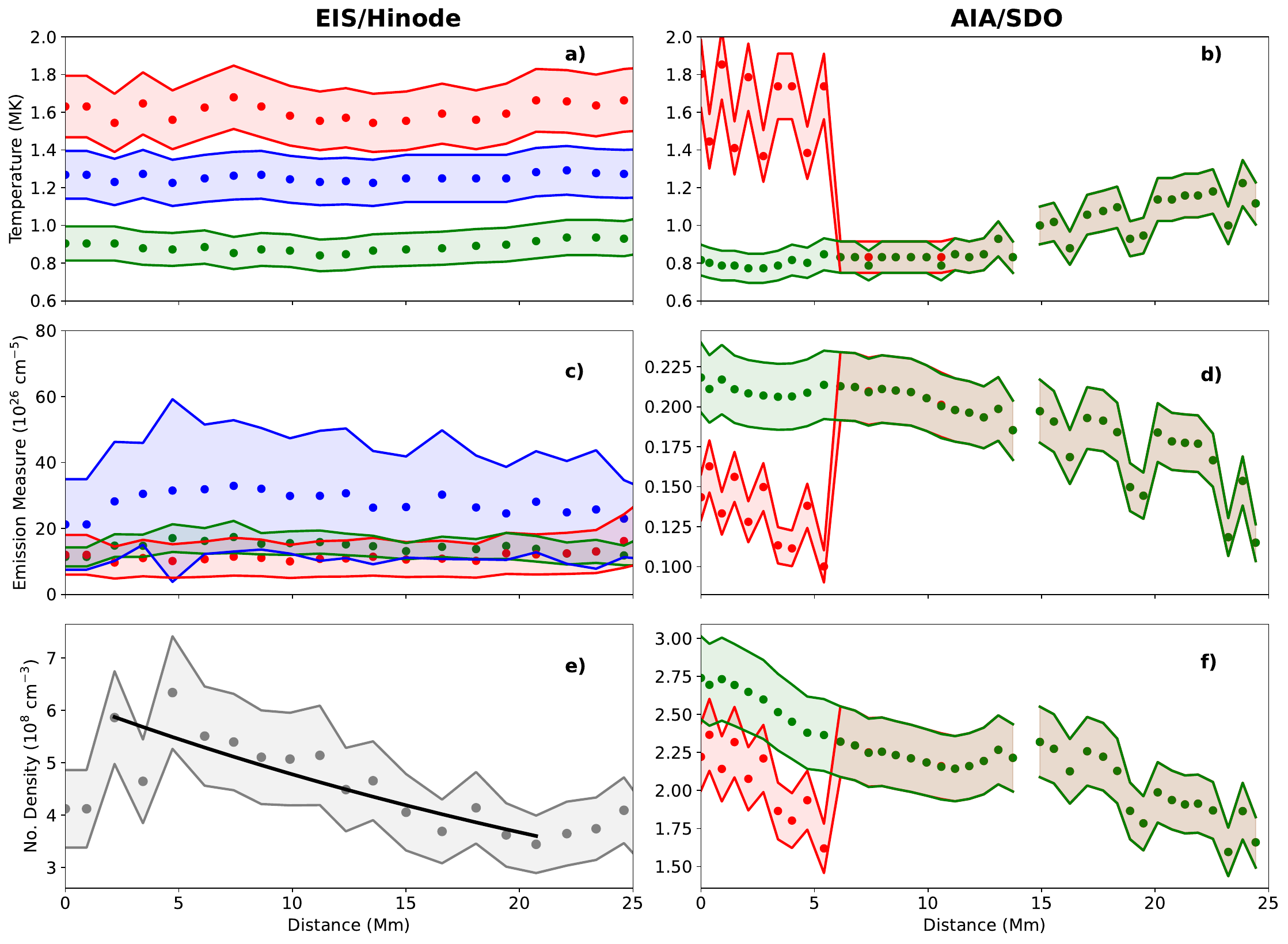}
    \caption{(a) Temperature obtained along the coronal loop from the EM-loci curves. The red, blue, and green colors represent high, average, and low temperature variation along the loops. (b) Temperature variation along the loop obtained from DEM analysis. Red and green colors represent the temperatures obtained from the high- and low-temperature DEM peaks. Overlapping region represents the temperature obtained from the single DEM peak. (c) Average EM variation along the coronal loop as obtained from the EM loci curves for three temperatures, shown in panel a. (d) Average EM variation along the coronal loop as obtained from the DEM analysis for three temperatures, shown in panel b. (e) Electron number density obtained from the Fe XII line ratio method of spectroscopic data. (f) Electron number density obtained from the DEM method of imaging data.}
    \label{fig:plasma}
\end{figure*}

To determine the average temperature along the loop, we plotted a histogram of the frequency of EM loci intersections as a function of temperature. In \autoref{fig:histo}, the x-axis represents temperature, and the left y-axis represents the frequency of intersections at that particular temperature. The histogram exhibits three prominent peaks in the temperature ranges 0.7-1.0 MK, 1.0-1.5 MK, and 1.5-1.8 MK, consistent with the EM loci distributions shown in  \autoref{fig:em_loci}. These results further support the multi-thermal nature of the loops. To determine temperature variation along the loop in these temperature bands, we computed the average of the intersection points for spectral line combinations Fe VIII to Fe XI for the low temperature component, Fe XI to Fe XIV for the high temperature component, and Fe VIII to Fe XIV for the intermediate (average) temperature component. These combinations were selected based on the EM loci distributions presented in \autoref{fig:em_loci}. Fe XI contributes to both the low- and high-temperature ranges because its formation temperature lies between these regimes, as shown in \autoref{fig:cf}. The resulting low-, high-, and average-temperature profiles along the loops are shown in red, blue, and green, respectively, in \autoref{fig:plasma} (a). We will be following the same colors to represent these temperatures and their respective emission measures throughout this paper. The error bars on temperature are the standard deviations of the intersection points at each location along the loops.

Temperature estimates from EIS spectroscopic data with those obtained from AIA imaging data, we employ the DEM sparse inversion code developed by \citet{2015ApJ...807..143C}. The DEM analysis is performed using six  EUV coronal passbands of AIA 94, 131, 171, 193, 211, and 335 \AA. The overplotted black lines in \autoref{fig:histo} represent the EM curves obtained from DEM analysis at intervals of 5-pixels along the loop. In \autoref{fig:histo}, the x-axis represents the temperature, and the right y-axis represents the emission measure. The solid black EM curves exhibit a clear double-peaked temperature distribution, whereas the dashed EM curves show a single peak, despite the loop being spatially resolved. The overplotted green dash-dot lines represent the average EM profile across the entire loop and provide a good outline over the temperature inferred from the histogram.

The first DEM peak at the coronal footpoint is $\approx$0.85 MK, representing the emission from the fan loops \citep{2017ApJ...835..244G,2025ApJ...988L..26R}. However, due to the limited temperature resolution of the DEM inversion, the dual peaks are not resolved along the entire loop, resulting in a single peak at an intermediate temperature. To determine the temperature variation, we extracted the temperature at which the DEM curves peak. The first and second peaks depict the low and high temperature peaks, plotted in green and red colors, respectively. The overlapping colors represent the average temperature extracted from the single peak shown by overlapping region in \autoref{fig:plasma} (a). The temperature profile derived from the DEM analysis exhibits a slight increase along the loop, potentially reflecting LOS superposition of plasma at different temperatures, as shown in \autoref{fig:plasma} (b). The EM images obtained in these temperature ranges (see Appendix \autoref{fig:dem_img}) indicate that the relative contributions of the low and high temperature components vary along the loop.

\subsubsection{Emission measure}

 The emission measure (EM) variation across high, low, and average temperatures is shown in \autoref{fig:plasma} (c). The variation in EM exhibits similar trends to the intensity variations of the spectral lines associated with the respective temperature ranges, as shown in \autoref{fig:int}. We have also estimated EM along the loop from DEM curves shown in \autoref{fig:histo}. The emission measure from the first peak (low temperature) and second peak (high temperature) are plotted in green and red colors, respectively, while the overlapping data points represent the EM associated with the single-peaked DEM distributions, as shown in \autoref{fig:plasma} (d). EM curves derived from DEM analysis near the footpoint clearly show the presence of dual-temperature plasma components. The imaging diagnostics indicate that the low-temperature component has a higher EM than the high-temperature component, suggesting that cooler plasma dominates near the footpoint as shown in \autoref{fig:plasma} (d). 


\subsubsection{Density}

\begin{figure}
    \centering
    \includegraphics[width=0.95\linewidth]{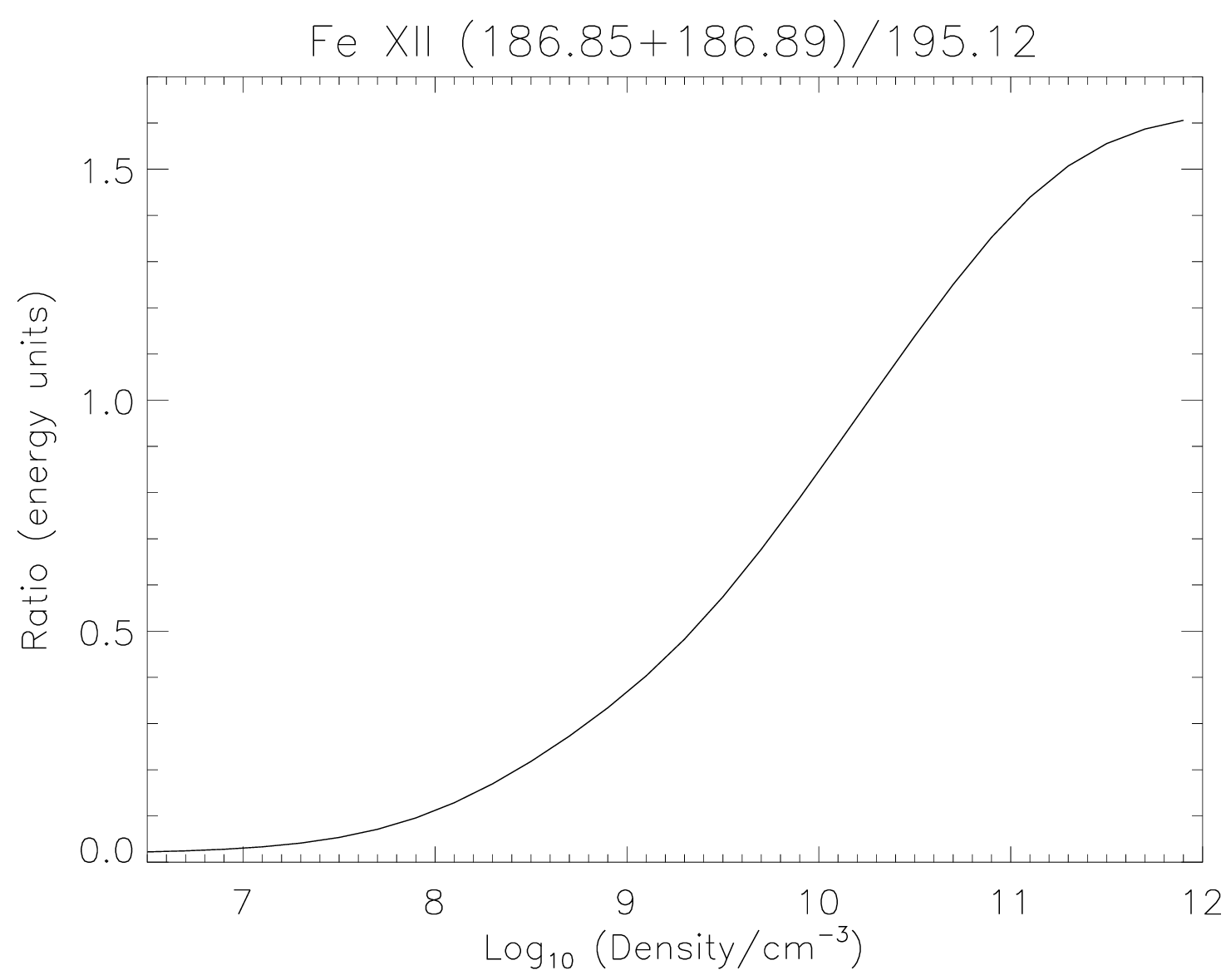}
    \caption{The intensity ratio of the electron density sensitive (Fe XII 186.88 \AA ) and insensitive (Fe XII 195.12 \AA ) spectral lines, as labeled, is obtained from CHIANTI v.10.1.}
    \label{fig:rat}
\end{figure}

Electron number density ($N_e$) is derived by using the line ratio method, which utilizes the intensity ratio of two emission lines of the same ion with different electron density dependence \citep[e.g.,][]{2007PASJ...59S.857Y,2008uxss.book.....P}. The line-ratio method provides the mean electron number density along the LOS. The variation of intensity ratio of the two spectral lines, with one sensitive and the other insensitive to the electron density, is obtained from CHIANTI v.10.1. Here we utilize the ratio obtained from density-sensitive line pairs of Fe XII 186.88/195.12 \AA\, as shown in \autoref{fig:rat}. The ratio is sensitive to changes between $10^8-10^{11.5}$ cm$^{-3}$, which is well within the coronal density range as shown in \autoref{fig:rat}. Using this ratio, we determine the electron number densities along the loop length as shown in \autoref{fig:plasma} (e). 

We fitted the electron number density variation along the coronal loop with an exponential decay function $N_e= N_o$exp$(-h/H_d)$ shown by the overplotted black line in the \autoref{fig:plasma} (e). The fitting is performed using MPFIT routines \citep{2009ASPC..411..251M}. The fit provides the density scale height $H_{d} \approx$ 38.0$\pm$10.4 Mm. We further estimated the electron number density ($N_e$) along the loop using AIA EM values at peak temperatures as shown in \autoref{fig:plasma} (f), using 
\begin{equation}
    N_e = \sqrt{\frac{EM_{col}}{w}},
\end{equation} where w is the diameter of the loop. Here, we assumed a fully ionized plasma in the corona and the loop filling factor to be 1.

\subsubsection{Velocity amplitude}

\begin{figure*}
    \centering
    \includegraphics[width=0.95\linewidth]{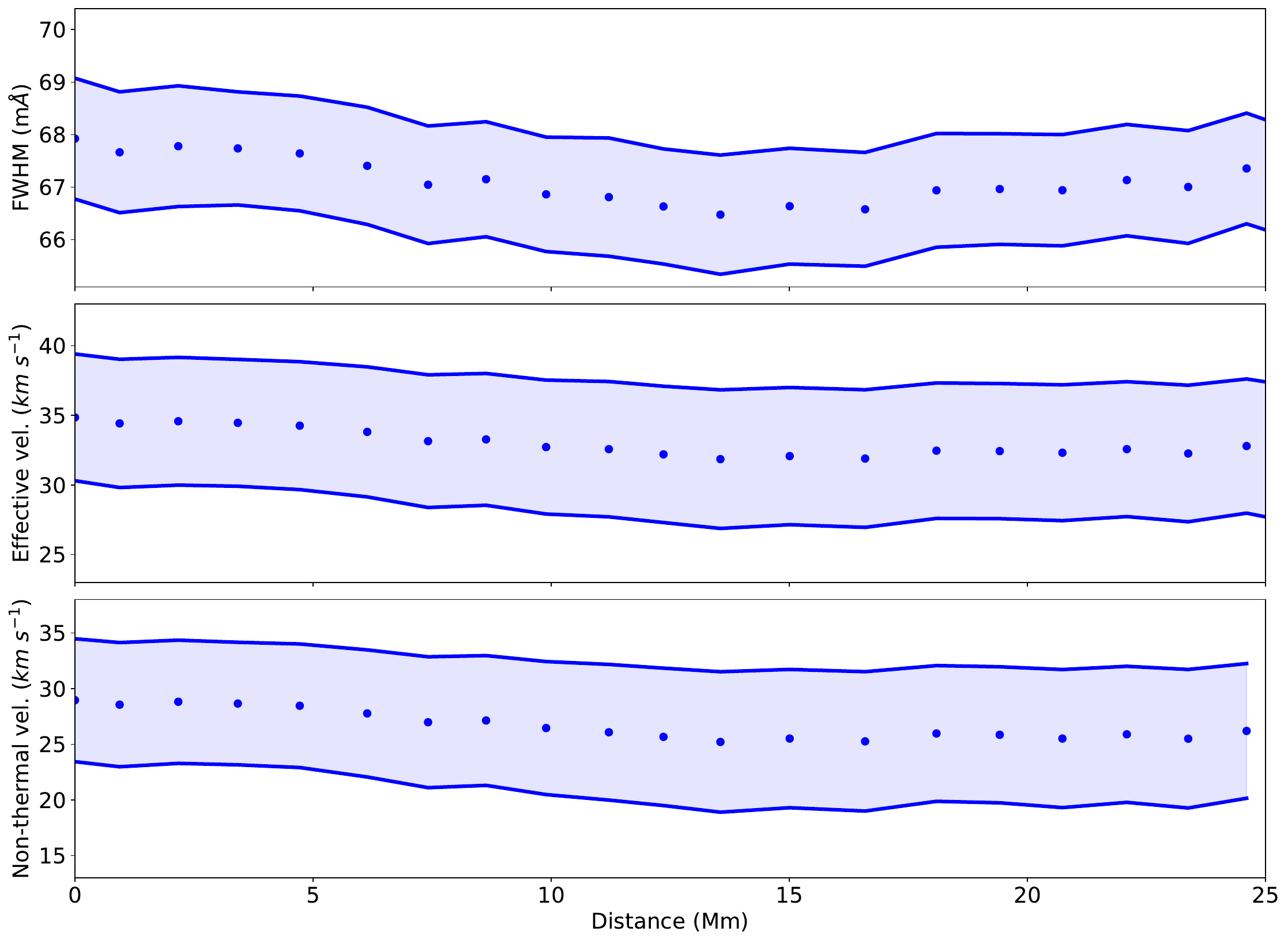}
    \caption{Variation of FWHM (top panel),  effective velocity (middle panel), and non-thermal velocity (bottom panel) obtained along the coronal loop from the Fe XII 195.12 \AA\ spectral line.}
    \label{fig:vel}
\end{figure*}

We further obtained the FWHM of the spectral line by fitting a Gaussian profile over the observed intensity profiles of spectral lines, as shown in \autoref{fig:prof}. FWHM is given by 
\begin{equation}
    FWHM= \sqrt{[4ln2 (\frac{\lambda}{c})^2 (\frac{2k_bT_i}{M_i}+v_{nth}^2)+W_{inst}^2]},
\end{equation}

where $k_b$ is Boltzmann constant, $\lambda$ is wavelength of spectral line, $c$ is speed of light, $T_i$ is ion temperature, $M_i$ is ion mass, $v_{nth}$ is non-thermal velocity, and $W_{inst}$ is the instrumental width. Variation of FWHM along the coronal fan loop is shown in the top panel of \autoref{fig:vel}. The EIS instrumental width for 1" slit varies between 56 to 59 m\AA\ for a downloaded central 512 pixels is calculated using IDL routine EIS$\_$SLIT$\_$WIDTH \footnote{https://sohoftp.nascom.nasa.gov/solarsoft/hinode/eis/doc/eis\_notes/07\_LINE\_WIDTH/eis\_swnote\_07.pdf}.

After subtracting instrumental width and converting FWHM to velocity using 
\begin{equation}
   \frac{\Delta \lambda}{\lambda}= \frac{v}{c}.
\end{equation}
we determined the effective velocity $v_{eff}$ as

\begin{equation}
    v_{eff}= \sqrt{\frac{2k_bT_i}{M_i}+v_{nth}^2}.
\end{equation}

The effective velocity obtained from Fe XII 195.12 \AA\, which is the strongest spectral line in our analysis, is plotted in middle panel of \autoref{fig:vel}. The error bar on effective velocities is calculated using error propagation from the fitting profile and instrumental width error of $\approx$2 m\AA.

\subsection{Alfv\'en wave energy flux}

Non-thermal velocity amplitude ($v_{nth}$) is an important parameter in calculating Alfv\'en wave energy flux and is assumed to represent the Alfv\'en wave velocity amplitude. The non-thermal velocity is calculated after subtracting the thermal velocity (broadening) contribution from the effective velocity. The thermal component is calculated assuming the plasma is in thermal equilibrium; therefore, we have approximated $T_e = T_i$. The resulting non-thermal velocity amplitude is shown in the bottom panel of \autoref{fig:vel} for the average temperature component. The associated errors were estimated by propagating the errors in the effective velocity and temperature. The non-thermal velocity amplitude of these waves is almost constant within the error range.

Alfv\'en wave energy flux is calculated using
\begin{equation}
    E_A= \frac{1}{2}\rho v_{nth}^2 v_A= \frac{1}{2}\sqrt{\frac{\mu m_p N_e}{4\pi}}v_{nth}^2B,
\end{equation}

where $v_A= \frac{B}{\sqrt{4\pi\rho}}$ is Alfv\'en speed, $B$ is magnetic field and $\rho$ is mass density ($\rho= \mu m_p N_e$, $m_p$ is proton mass, $\mu$=1.27 is mean molecular mass),. Here, we assume an average magnetic field strength of 200 G inferred at the coronal loop footpoint \citep[see][Table 1]{2025ApJ...988L..26R}. The calculated Alfv\'en wave energy flux is $\approx 2 \times 10^7$ erg cm$^{-2}$ s$^{-1}$. 

We performed a similar analysis on the other five loops identified in \autoref{fig:fan loop}. We found that all the loops are multi-thermal and overdense. Here also, Alfv\'en energy flux at coronal loop footpoint is $\approx 1-2 \times 10^7$ erg cm$^{-2}$ s$^{-1}$ as found for loop 1. The estimated plasma and wave parameters at all the loop footpoints are provided in \autoref{tab:stat}. 

\begin{table*}
    \caption{Plasma and wave parameters derived at coronal loop footpoints using EIS spectroscopic data.}
     \label{tab:stat}
   \centering
   \small
       \resizebox{0.8\textwidth}{!}{
    \begin{tabular}{|c|c|c|c|c|c|c|c|c|c|c|} 
    \hline
\multicolumn{1}{|c|}{Loop}   &  
\multicolumn{3}{|c|}{ Temperature }  &
\multicolumn{1}{|c|}{ Density }  &
\multicolumn{1}{|c|}{ $v_{nth}$ velocity }  &
\multicolumn{1}{|c|}{Alfv\'en wave flux}  \\ 

 no.  & MK & MK & MK &cm$^{-3}$ & km s$^{-1}$ & erg cm$^{-2}$s$^{-1}$ \\ \hline
1  & 0.84 & 1.24& 1.61  & 5.10$\times$10$^8$ &26.56 &1.31$\times$10$^{7}$\\ \hline
2  & 0.85& 1.26 & 1.70 & 4.82$\times$10$^8$ & 27.16 &1.33$\times$10$^{7}$ \\ \hline
3  & 0.91& 1.35 & 1.75 & 5.79$\times$10$^8$ & 30.36 &1.82$\times$10$^{7}$ \\ \hline
4  & 0.94& 1.33 & 1.77 & 4.80$\times$10$^8$ & 29.64 &1.58$\times$10$^{7}$ \\ \hline
5  & 0.89& 1.26 & 1.59 & 6.53$\times$10$^8$ & 30.31 &1.93$\times$10$^{7}$ \\ \hline
6  & 0.78& 1.30 & 1.76 & 9.66$\times$10$^8$ & 26.25 &1.76$\times$10$^{7}$ \\ \hline
    \end{tabular}}
\end{table*}


\section{Discussion and conclusion}
\label{sec:summary}

In this work, we selected coronal fan loops that are clearly visible in both the hot- and cool-temperature passbands of AIA and spectral lines of EIS. Using EIS spectroscopic and AIA imaging data, we estimated various plasma parameters along coronal loops and compared the results obtained using different techniques. The loops were identified in multiple spectral lines covering the peak formation temperature of 0.5--2 MK, ideal for coronal fan loop studies. We determined the variation of the electron number density and temperature along the coronal loops using the line ratio and EM loci method, respectively, and compared the resulting variations with those obtained from DEM analysis of AIA observations.
 
 Results from the EM-loci technique clearly indicate that plasma along the LOS is multi-thermal, as all EM curves intersect at three different points. These three different intersection regions (temperatures) are present along the whole loop. The temperature histogram derived from the intersections reveals three prominent components, with temperatures of approximately 0.8–0.9 MK, 1.2–1.4 MK, and 1.6–1.8 MK. These components persist along the observed loop structures, with a similar temperature range present in all the analyzed loops (see \autoref{tab:stat}).

 DEM temperature and emission measure clearly show that low temperatures dominate near the loop footpoints, while temperatures gradually increase farther along the loops. The clear, persistent multi-thermal nature of these fan loops is also visible from the spectral analysis. Despite this multi-thermal structure, the temperature profiles of the individual components remain approximately constant within the error bars, highlighting that each temperature component is isothermal along the loop. Similar isothermal behavior has been reported by \citet{2019A&A...627A..62G} along long coronal loop. The observed multi-thermal nature may be due to unresolved, multi-stranded loops that cannot be spatially resolved by EIS and AIA. However, this can be sufficient for measuring reasonable non-thermal Doppler broadening measurements at larger scales along these loops \citep[e.g.,][]{2026PASJ...78.1524M}. Future observations with the Multi-slit Solar Explorer \citep[MUSE;][]{2022ApJ...926...52D} and EUV High-Throughput Spectroscopic Telescope \citep[EUVST;][]{2026arXiv260910997T} onboard SOLAR-C, can resolve the fine-scale structure of these loops and clarify the relationship between their thermal structure and wave dynamics within the loops.
 
 The plasma temperature near the footpoints may be lower than the values estimated in our analysis. The fan loop footpoint region is also associated with bright emission observed in the Mg V 2796.16 \AA\ line probing 0.32 MK (see \autoref{appendix:footpoint}) \citep{2007PASJ...59S.857Y}. We also identify high temperature contribution in AIA 304 \AA\ passband, as shown in \autoref{appendix:dem_304} \citep[e.g.,][]{2024SoPh..299...94A}. However, this needs further detailed investigation. These observations suggest that the thermal structure near the footpoints may be more complex than those inferred from the primary diagnostics used here. However, further investigation is required to establish the origin and significance of these contributions.

The comparison between spectroscopic temperatures and those derived from AIA DEM analysis also supports the multi-thermal nature of the loops. The DEM profiles resolve two distinct temperature components only over a limited segment of approximately 5 Mm. Farther along the loops, the two peaks merge into a single broad peak, yielding an intermediate temperature. This limitation may partly arise from the broad temperature response functions of the AIA passbands, which restrict the ability of imaging-based DEM inversion to distinguish closely spaced temperature components. In contrast, the spectroscopic EM loci method reveals multiple temperature components along the longer length of the observed loops. The comparison demonstrates the complementary capabilities of spectroscopic and imaging diagnostics in characterizing the thermal structure of coronal loops. The intermediate temperature component 1.2-1.4 MK identified in the spectroscopic observations is not evident in the emission measure (EM) distribution obtained from AIA imaging data. A noticeable dip is observed at the corresponding temperature range, as shown in \autoref{fig:histo}. 
A further detailed investigation is required to establish the underlying cause of this discrepancy.

The emission measure variation along the loop further shows that the low-temperature component dominates near the footpoints, where its EM exceeds that of the high-temperature component. This dominance persists over the initial segment of loop approximately upto 5 Mm. Farther along the loops, the relative contributions of the two components merge, and we get a single intermediate EM. Beyond this region, the DEM analysis no longer consistently resolves the two components separately. These results indicate that the relative contributions of the temperature components vary along the loops, highlighting the importance of accounting for multi-thermal plasma when interpreting coronal observations. 

We obtained the electron density along the loop using the density-sensitive EIS spectral line pair Fe XII $\lambda$ 186.88 and $\lambda$ 195.12 \AA. Obtained number densities from spectroscopic data is around 6$\times$ 10$^8$ cm$^{-3}$ near the footpoint, which falls to $\approx$ 4$\times$ 10$^8$ cm$^{-3}$ farther along the loop. The expected density scale height from hydrostatic equilibrium at T$_e\approx$1.24 MK is $\lambda(T_e)\approx 46[\frac{T_e}{1MK}]$ Mm $\approx$ 57 Mm \citep{1999ApJ...520..880A,2019A&A...627A..62G}. The obtained POS density scale height $38.0\pm 10.4$ Mm is less than the expected density scale height. The difference in scale heights can be either due to the inclination of the loop \citep[e.g.,][]{2025ApJ...988L..26R}, or the overdense nature of loops, or both.
 \citet{2003ApJ...593.1164W} reported that long cool coronal loops are overdense, whereas short hot loops are underdense when compared with the static equilibrium solutions. Such departures from static models may be associated with impulsive heating models \citep[e.g.,][see details in \autoref{appendix:dem}]{2008ApJ...682.1351K}, along with other diagnostic techniques to constrain the heating mechanisms operating in the analyzed fan loops. Moreover, recent simulation results of \citet{2021ApJ...922L..26D} suggest that Alfv\'en waves can also induce overdense threads in the solar Corona. 

We also determined the total, effective, and non-thermal velocity profiles along the loops.  The effective velocity is derived from the full width at half maximum (FWHM) of Gaussian fits to the observed spectral line profiles after subtracting the instrumental broadening. The thermal velocity (broadening) is estimated by assuming electron–ion temperature equilibrium, and the non-thermal velocity (Alfv\'en wave velocity amplitude) is obtained by removing the thermal contribution from the effective velocity. The strongest line in our analysis Fe XII 195.12 \AA\ shows an almost constant non-thermal velocity along the loop length, and is $\approx$25-35 km s$^{-1}$, similar to those noted in earlier reports  \citep[e.g.,][]{2016ApJ...820...63B,2019A&A...627A..62G}. The near constant non-thermal velocity may provide constraints on the evolution of Alfv\'en wave perturbations and their energy transport along the loops.

 Upon combining all the measured parameters with the magnetic field strength of $\approx200$ G at the coronal footpoint of the loop \citep{2021ApJ...915L..24B,2025ApJ...988L..26R}, we obtained Alfv\'en wave energy flux of $\approx 10^7$ erg cm$^{-2}$s$^{-1}$. This energy flux is sufficient to maintain the hot active region corona \citep{1977ARA&A..15..363W}. Furthermore, time–distance analysis along fan loops from AIA/SDO 171 \AA\ imaging observations also reveals propagating SMAWs along the same loop structures (see \autoref{appendix:td}). Unclean ridges observed here may result from superposition of SMAWs propagating along different temperature strands. These multi-thermal loops can be an excellent site to explore multi-thermal apparent damping \citep[MAD;][]{2024A&A...683A.109V}.
 
 Future work with MUSE and EUVST/SOLAR-C will focus on (i) combining the EM-loci histogram technique with DEM analysis to investigate the origin of dips and unresolved temperature components in DEM distributions; (ii) exploring effects of wave superposition in complex, multi-thermal coronal structures on the wave properties and energy transport; and (iii) assessing the influence of background subtraction on the derived plasma parameters and wave energy estimates. These investigations will help refine our understanding of the thermal structure, wave dynamics, and energy budget of active-region coronal loops.

\section*{Acknowledgments}
The research work at the Physical Research Laboratory (PRL) is funded by the Department of Space, Government of India. AR thanks PRL for her post-doctoral research fellowship. AIA data are courtesy of SDO (NASA). Hinode is a Japanese mission developed and launched by ISAS/JAXA, with NAOJ as domestic partner and NASA and STFC (UK) as international partners. It is operated by these agencies in co-operation with ESA and NSC (Norway). CHIANTI is a collaborative project involving George Mason University, the University of Michigan (USA), University of Cambridge (UK) and NASA Goddard Space Flight Center (USA).

\section*{Data Availability}

The observational data utilized in this study from AIA on-board SDO are available at \url{http://jsoc.stanford.edu/ajax/lookdata.html} and data from Hinode mission are available at \url{https://sdc.uio.no/search/form}.

\appendix


\section{Loop footpoint temperature}
\label{appendix:footpoint}

\begin{figure*}
    \centering
    \includegraphics[width=0.9\linewidth]{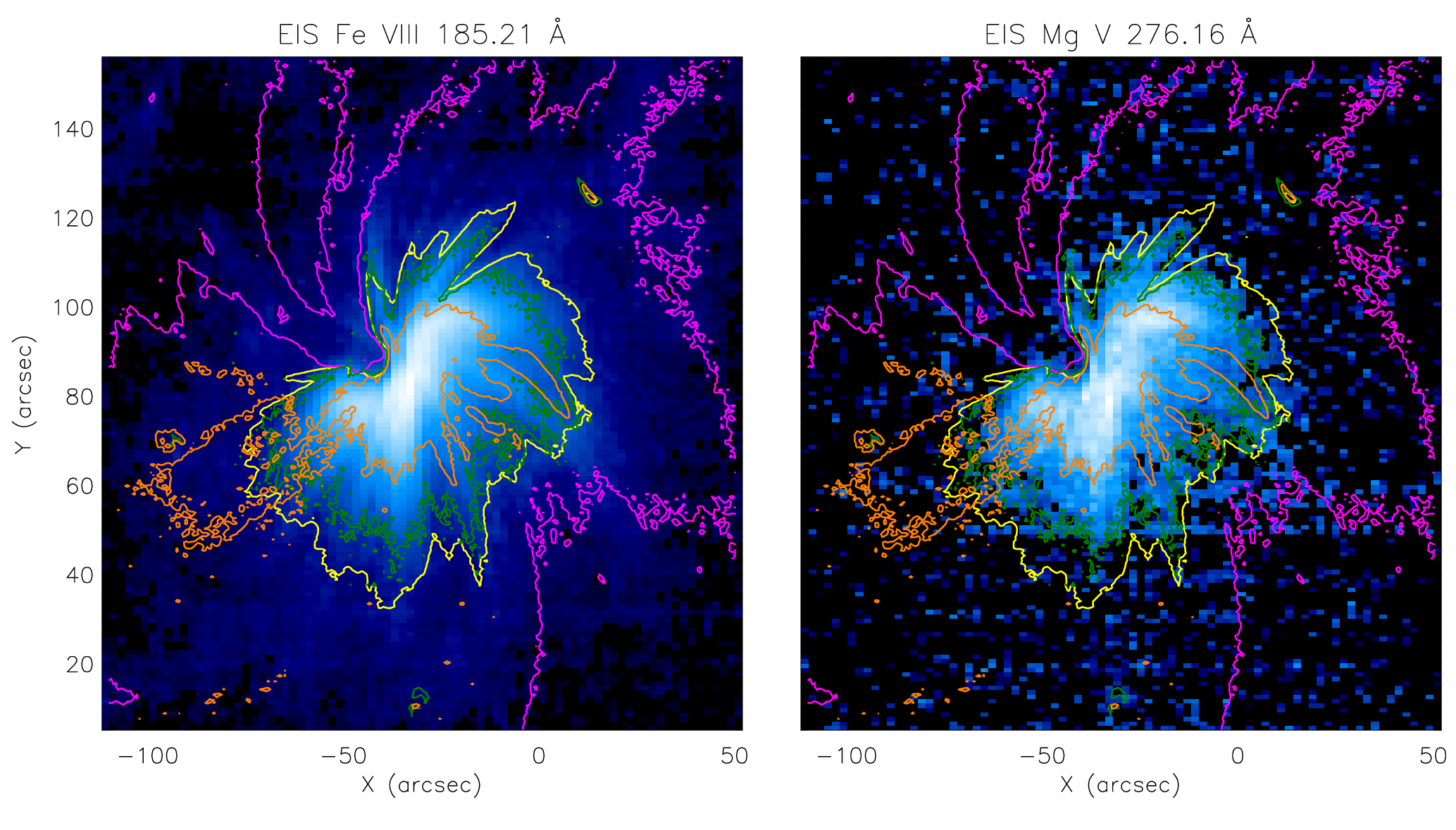}
    \caption{Monochromatic intensity maps of fan loop system obtained from low temperature EIS spectral lines as labeled. Here, the overplotted yellow, pink, green, and orange contours correspond to AIA 131, 171, 193, and 211 \AA\ passbands, respectively. }
    \label{fig:eis_mg}
\end{figure*}

In left panel of \autoref{fig:eis_mg}, we show EIS Fe VIII 185.21 \AA\ intensity image. The intensity contours extracted from AIA 131, 171, 193, and 211 \AA\ passbands, are plotted in yellow, pink, green, and orange. These intensity contours accurately map over the EIS Fe VIII image, demonstrating the alignment accuracy between EIS images with AIA passbands.

According to \citet{2007PASJ...59S.857Y}, the Mg V line at 276.16 \AA\ with peak formation temperature of 0.32 MK (transition region temperature) is very weak in most circumstances. However, the line is significantly enhanced at loop footpoints as shown in the right panel of \autoref{fig:eis_mg}. The intensity contours extracted from AIA 131, 171, 193, and 211 \AA\ passbands, plotted in yellow, pink, green, and orange maps, show the high-intensity regions over Mg V image, demonstrating strong transition region contribution in these passbands. This further highlights the low temperature at the coronal footpoint, and can be a topic of future exploration. 

\section{Loop stability}
\label{appendix:td}
\begin{figure*}
    \centering
    \includegraphics[width=0.8\linewidth]{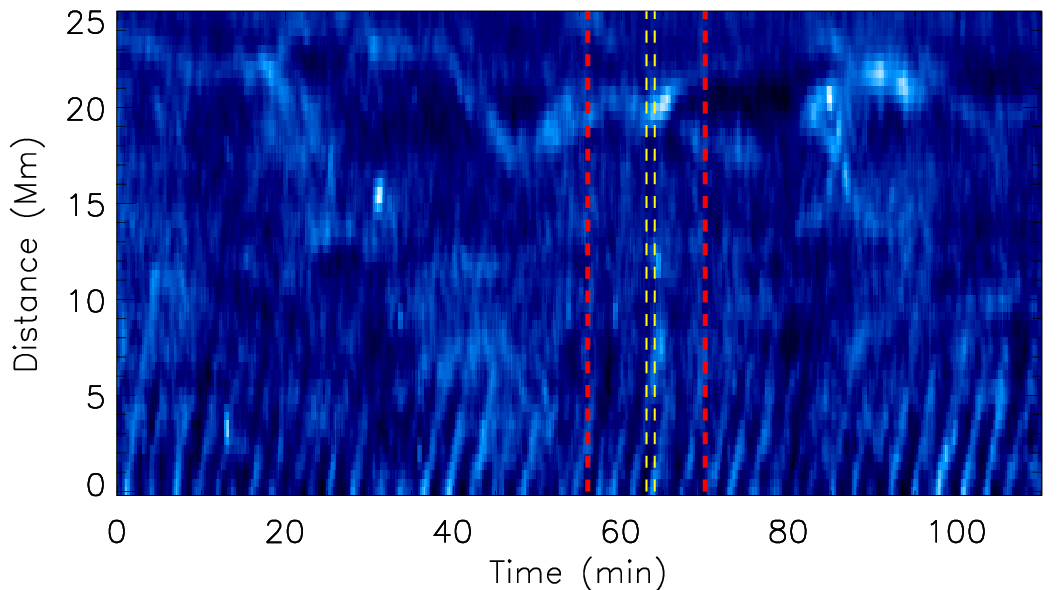}
    \caption{Time–distance maps produced from detrended light curves along Loop 1 from AIA 171 \AA\ passband. The 3-min SMAWs propagating along the loop appears as alternate bright and dark ridges. The red dashed box shows the time range when the loop was rastered by EIS. The yellow box shows the 1-minute time over which the AIA intensities are averaged to perform DEM analysis.}
    \label{fig:td}
\end{figure*}

To determine the stability of the loop with time when the loop was rastered by EIS, we obtained the detrended light curves by subtracting the background light curves (18-min running average of the original light curves) from the original light curves at those loop locations from AIA 171 \AA\ passband. These detrended light curves are utilized to create time-distance intensity map along the fan loop, as shown in \autoref{fig:td}. The time-distance image shows intensity evolution over time, with alternating bright and dark ridges, indicating the presence of propagating 3-min SMAWs along the fan loop \citep[e.g.,][]{2020A&A...638A...6S}. The time-distance plot is showing unclean ridges compared to the previous observations \citep[e.g.,][]{2026ApJR}. These unclean ridges can be due to the superposition of different waves propagating along different temperature strands. These multi-thermal loops can be an excellent site to explore  multi-thermal apparent damping of slow waves \citep[MAD,][]{2024A&A...683A.109V}. Here, we are plotting a 2-hour time series during which EIS observed the fan-loop system. The red box indicates the time interval during which the loop was rastered by EIS. The yellow box shows the 1-minute time over which the intensities were averaged to perform DEM analysis using AIA observations. We found that there are no strong brightening within a first few Mm of the coronal loop footpoint region. Fan loops did not evolve when the slit rastered over it, indicating that these are quiescent loops.

\section{Fan loop in different temperature ranges}
\label{appendix:dem}
\begin{figure*}
    \centering
    \includegraphics[width=\linewidth]{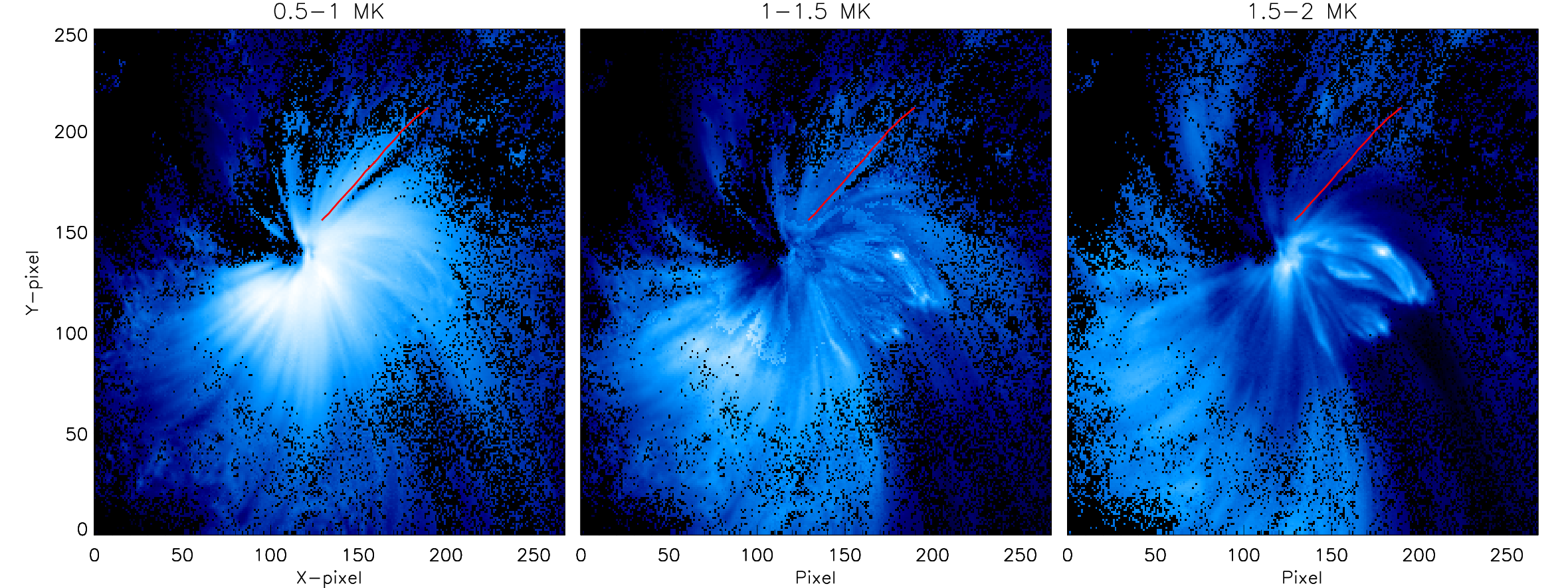}
    \caption{Emission measure (EM) images obtained in different temperature bins as labeled. Images highlight the temperature variation along the loops.}
    \label{fig:dem_img}
\end{figure*}

To further explore the multi-thermal nature, we created EM images of the fan loop system in different temperature ranges, 0.5-0.9 MK, 1.1-1.5 MK, and 1.5-2 MK using DEM tool provided by \citet{2015ApJ...807..143C}. Analysis is performed by averaging intensities over 1-minute at the time when EIS was rastering Loop 1 under the study, as shown by the yellow box in \autoref{fig:td}. The loop is visible over the entire temperature range; however, we observe a change in its intensity and structure as we move to higher temperatures. In the left panel of \autoref{fig:dem_img}, i.e., in the low temperature range, the footpoint region of the loops are bright. This footpoint temperature can be much lower than what is calculated here with the AIA DEM analysis, as the loop footpoints are also visible in Mg V image created with EIS which the probes the transition region temperature of 0.3 MK, as described in \autoref{appendix:footpoint}.  In the middle panel of \autoref{fig:dem_img}, i.e., average temperature range, the footpoint region of the loops are slightly dark. In the right panel of \autoref{fig:dem_img}, i.e., in the high temperature range, the middle portion of the loops are dark, and the footpoint region of the loops are bright, which can be related to impulsive heating model \citep{2008ApJ...682.1351K}. Additionally, the structure appears to be elongated in the high temperature range. These highlight the multi-thermal temperature variation along the loops.

\section{Hot temperature in the AIA 304 \AA\ passband}
\label{appendix:dem_304}

\begin{figure*}
    \centering
    \includegraphics[width=0.6\linewidth]{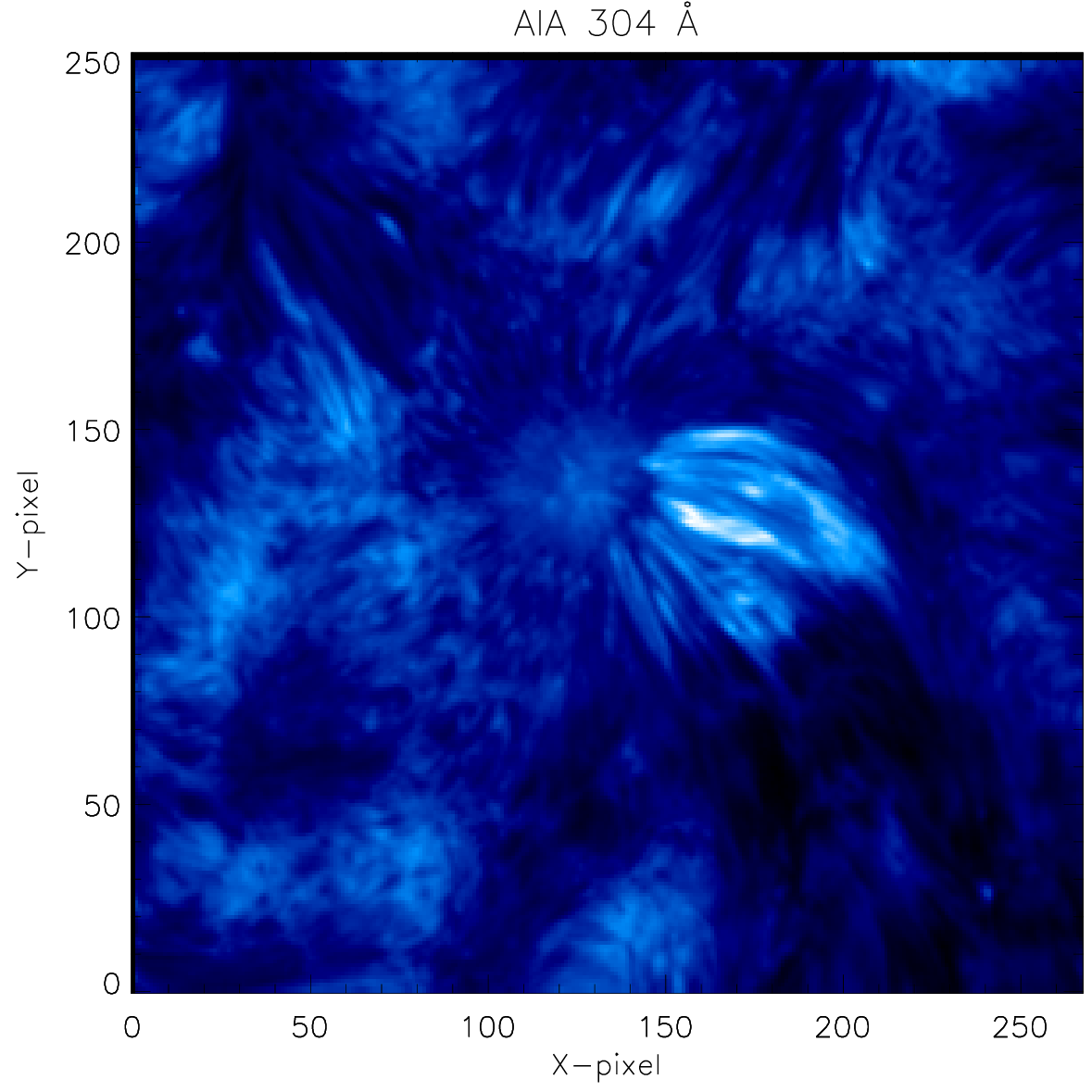}
    \caption{High temperature contribution in AIA 304 \AA\ passband. }
    \label{fig:aia304}
\end{figure*}

The DEM image shown in the hot temperature range 1.5-2 MK, in \autoref{fig:dem_img}, exhibits a similar structure in the center left portion of the EM image to that of the AIA 304 \AA\ passband image shown in \autoref{fig:aia304}. This morphological similarity suggests that the AIA 304 \AA\ passband contains a coronal emission contribution from Si XI emission at 303.32 \AA\ forming at around 1.5 MK, as also noted by \citet{2026SoPh..301...70A}.

\label{lastpage}

\end{document}